\documentclass[aps,twocolumn,pra,superscript,floatfix,superscriptaddress,showpacs]{revtex4-1}

\usepackage{graphicx}% Include figure files
\usepackage{dcolumn}% Align table columns on decimal point
\usepackage{bm}% bold math
\usepackage{color,soul}

\usepackage{amsmath}
\usepackage{amssymb} %DIF > 
\usepackage{mathrsfs} %DIF > 
\usepackage[colorlinks=true,%
            linkcolor=blue,%
            urlcolor=blue,%
            citecolor=blue,%
            filecolor=blue,%
            bookmarksopen=true,%
            pdfauthor={AGarrido},%
            pdftitle={QDTcrossThermoelectricMZM},%
            pdfsubject={QDTcrossThermoelectricMZM_Manuscript},%
            pdfpagemode=UseOutlines]{hyperref}

\definecolor{applegreen}{rgb}{0.55, 0.71, 0.0}
\definecolor{antiquefuchsia}{rgb}{0.57, 0.36, 0.51}
\definecolor{amethyst}{rgb}{0.6, 0.4, 0.8}

\RequirePackage{color} %DIF PREAMBLE
\providecommand{\DIFaddtex}[1]{{\protect\color{blue}#1}} %DIF PREAMBLE
\providecommand{\DIFdeltex}[1]{} %DIF PREAMBLE
\providecommand{\DIFaddbegin}{} %DIF PREAMBLE
\providecommand{\DIFaddend}{} %DIF PREAMBLE
\providecommand{\DIFdelbegin}{} %DIF PREAMBLE
\providecommand{\DIFdelend}{} %DIF PREAMBLE
\providecommand{\DIFdelFL}[1]{} %DIF PREAMBLE
\providecommand{\DIFadd}[1]{\texorpdfstring{\DIFaddtex{#1}}{#1}} %DIF PREAMBLE
\providecommand{\DIFdel}[1]{\texorpdfstring{\DIFdeltex{#1}}{}} %DIF PREAMBLE
\RequirePackage{listings} %DIF PREAMBLE
\RequirePackage{color} %DIF PREAMBLE
\lstdefinelanguage{DIFcode}{ %DIF PREAMBLE
  moredelim=[il][\color{red}\scriptsize]{\%DIF\ <\ }, %DIF PREAMBLE
  moredelim=[il][\color{blue}\sffamily]{\%DIF\ >\ } %DIF PREAMBLE
} %DIF PREAMBLE
\lstdefinestyle{DIFverbatimstyle}{ %DIF PREAMBLE
	language=DIFcode, %DIF PREAMBLE
	basicstyle=\ttfamily, %DIF PREAMBLE
	columns=fullflexible, %DIF PREAMBLE
	keepspaces=true %DIF PREAMBLE
} %DIF PREAMBLE
\lstnewenvironment{DIFverbatim}{\lstset{style=DIFverbatimstyle}}{} %DIF PREAMBLE
\lstnewenvironment{DIFverbatim*}{\lstset{style=DIFverbatimstyle,showspaces=true}}{} %DIF PREAMBLE
\begin{document}

%\preprint{AAPM/123-QED}

\title{Phase-controlled quasi-bound states in the continuum and thermoelectric enhancement in Majorana–quantum-dot nanostructures}

\author{A. P. Garrido}
\email{alejandro.garridoh@usm.cl}
\email{a.garridohidalgo@uandresbello.edu}

\affiliation{ 
Departamento de F\'isica, Universidad T\'ecnica Federico Santa Mar\'ia, Av. Espa\~na 1680, Casilla 110V, Valparaiso, Chile.%\\This line break forced with \textbackslash\textbackslash
}%
\affiliation{ 
Departamento de Física y Astronomía, Facultad de Ciencias Exactas, Universidad Andres Bello, Sazié 2212, Santiago, Chile.%\\This line break forced with \textbackslash\textbackslash
}%

\author{D. Zambrano}
\affiliation{ 
Departamento de Física, Facultad de Ciencias Naturales, Matemática y del Medio Ambiente,
Universidad Tecnológica Metropolitana,
Las Palmeras 3360, Ñuñoa 780-0003, Santiago, Chile.%\\This line break forced with \textbackslash\textbackslash
}%

\author{H. Farfán-Bachiloglu}
\affiliation{Instituto de Física, Pontificia Universidad Católica de Valparaíso,
Avenida Universidad 331, Curauma, Valparaíso, Chile.}
\author{J. P. Ramos-Andrade}
\affiliation{Departamento de F\'isica, Universidad de Antofagasta, Av. Angamos 601, Casilla 170, Antofagasta, Chile.}

\author{Vladimir Juri\v{c}i\'c}%
 \affiliation{ 
Departamento de F\'isica, Universidad T\'ecnica Federico Santa Mar\'ia, Av. Espa\~na 1680, Casilla 110V, Valparaiso, Chile.%\\This line break forced with \textbackslash\textbackslash
}%

\author{P. A. Orellana}%
 \affiliation{ 
Departamento de F\'isica, Universidad T\'ecnica Federico Santa Mar\'ia, Av. Espa\~na 1680, Casilla 110V, Valparaiso, Chile.%\\This line break forced with \textbackslash\textbackslash
}%

\date{\today}% It is always \today, today,

\begin{abstract}
We investigate how the interplay between Majorana zero modes (MZMs) and bound states in the continuum (BICs) governs the electronic  thermoelectric response of a crossbar-shaped quantum dot (QD) coupled to two topological-superconductor nanowires.
Using the Green-function formalism, exact linear-response energy integrals, and their low-temperature Sommerfeld expansion, we analyze the spectral and thermoelectric properties of the system.
We show that symmetry breaking converts BICs into quasi-BICs, allowing them to contribute to electrical and thermal transport and thereby generate a finite thermoelectric response.
While unequal nanowire lengths, reflected in different intra-Majorana coupling strengths, produce only a modest enhancement of $ZT_{\rm el}$, detuning the QD level increases $ZT_{\rm el}$ by approximately one order of magnitude.
Superconducting-phase control produces a much stronger enhancement, reaching $ZT_{\rm el}\simeq0.75$ through a quadratic transmission zero and 
a pronounced violation of the Wiedemann--Franz law.
The low-temperature values $ZT_{\rm el}^{\max}\simeq0.755$ and $\mathscr{L}/\mathscr{L}_0=21/5$ are universal consequences of this quadratic antiresonance. Our results establish phase-tunable thermoelectric signatures of the Majorana-coupled interference structure and identify superconducting-phase control as an efficient means of engineering the electronic response of topological hybrid nanostructures.
\end{abstract}

\DIFdelbegin %DIFDELCMD < \keywords{Majorana zero modes, quantum dots, Bound states in the continuum}%%%
\DIFdelend \DIFaddbegin \DIFaddend %Use showkeys class option if keyword.
\maketitle

%%%%%%%%%%%%%%%%%%%%%%%%%%%%%%%%%%%%%%%%%%%%%%%%%%%%%%%%%%%%%%
\section{\label{sec:intro}Introduction}
%%%%%%%%%%%%%%%%%%%%%%%%%%%%%%%%%%%%%%%%%%%%%%%%%%%%%%%%%%%%%%

Topological superconductor nanowires (TSCNs) are at the forefront of condensed matter research, prized for their potential application in fault-tolerant quantum computing \cite{kitaev2003fault, nayak2008non, pachos2012introduction, beenakker2013search, laflamme2014publisher, albrecht2016exponential}. Exotic fermionic quasiparticles predicted within this framework, which behave as their own anti-quasiparticles, have emerged \cite{majorana1937teoria, wilczek2009majorana, franz2010race}. These quasiparticles, known as Majorana zero modes (MZMs)—localized states in topological superconductors—exhibit non-Abelian statistics, making their quantum state manipulable through braiding operations \cite{kraus2013majorana, alicea2011non}, establishing their potential for fault-tolerant quantum computation \cite{kitaev2001unpaired, bravyi2002fermionic, kitaev2003fault, nayak2008non, leijnse2011quantum, pachos2012introduction,  albrecht2016exponential}. These MZMs are predicted to localize at the ends of a TSCN, consisting of a semiconductor-superconductor nanowire with strong spin-orbit interaction, under a magnetic field. 
The aforementioned system can be viewed as a realization of a Kitaev chain \cite{kitaev2001unpaired, kitaev2003fault, moore2009next}, where the two MZMs located at opposite ends of the wire are coupled, with this coupling decaying exponentially with the wire's length \cite{albrecht2016exponential}. This decoupled MZMs behavior enables the construction of a qubit that is topologically protected against decoherence by local perturbations \cite{wu2012tunneling, kitaev2001unpaired, kraus2013majorana, albrecht2016exponential, semenoff2006teleportation, tewari2008testable}. Mourik and collaborators achieved the first physical realization of this system, reporting zero-bias anomalies in conductance as evidence of the presence of MZMs \cite{mourik2012signatures}. However, these anomalous signatures do not always provide conclusive evidence of MZMs, as they can often be mimicked by trivial effects such as disorder or Andreev bound states. This ambiguity underscores the pressing need for custom-designed experimental protocols capable of uniquely identifying the non-Abelian statistics that characterize true MZMs \cite{deng2012anomalous, mourik2012signatures, das2012zero, lee2012zero, finck2013anomalous, churchill2013superconductor, zambrano2018bound, ramos2019fano, garrido2023bound}. 
An alternative method to identify MZMs involves thermoelectric measurements, which offer distinct advantages by exploring their unique transport signatures. Although conventional thermoelectric measurement techniques were developed in the early 1990s \cite{Molenkamp1992:PRL,HvanHouten_1992:SST}, they have since evolved into potent tools for detecting chargeless MZMs. These techniques can disclose MZM signatures through thermal conductance \cite{Fu2008:PRL, Bauer2021:PRB,Ramos2016thermoelectric,garrido2025thermoelectric}, voltage thermopower \cite{Dolgirev2019:RRL,Hou2013:PRB,Sela2019:PRL,Ramos2016thermoelectric,garrido2025thermoelectric}, or the violation of the \DIFdelbegin \DIFdel{Wiedemann-Franz }\DIFdelend \DIFaddbegin \DIFadd{Wiedemann--Franz }\DIFaddend (WF) law \cite{Giuliano2022:PRB,Buccheri2022:PRB,Benjamin2024:EL,Ramos2016thermoelectric,garrido2025thermoelectric}, providing complementary evidence beyond zero-bias anomalies.

Recent experiments and device proposals have developed increasingly stringent local, nonlocal, and parity-sensitive protocols for identifying parameter regimes compatible with topological superconductivity \cite{pikulin2021protocol,c325-kgbf,aghaee2023inas,aasen2025roadmap,aghaee202620}. Nevertheless, unambiguous discrimination between MZMs and trivial subgap states remains an active challenge, motivating complementary probes that combine electrical, thermal, and interference-sensitive measurements.

Additionally, bound states in the continuum (BICs) do not decay even when their energy levels reside within the domain of continuum states \cite{hsu2016bound}. Originally predicted by von Neumann and Wigner at the dawn of quantum mechanics \cite{vonNeumann-Wigner}, BICs have garnered considerable interest, particularly following the observation of such states in photonic systems. Furthermore, owing to the similar interference phenomena observed in both electronic and photonic systems, the potential presence of BICs in electronic systems has been proposed \cite{ramos2014bound, hsu2016bound, zambrano2018bound, grez2022bound, garrido2023bound}. Thermoelectric efficiency studies have also revealed the characteristics of BICs, as well as information about their formation and stability \cite{Sierra:PRB2016,Ronald:SPPC2024,garrido2025thermoelectric}.

In this article, we study two topological-superconductor nanowires coupled in a crossbar geometry to a central quantum dot connected to two normal leads, as illustrated in Fig.~\ref{fig1}. We calculate the electrical and electronic thermal conductances, the dot and Majorana spectral functions, and the corresponding thermoelectric coefficients using the Green-function formalism. The transport moments are evaluated over the full energy axis, while the Sommerfeld expansion is used to expose the low-temperature structure. We compare three symmetry-breaking mechanisms: unequal Majorana hybridization energies caused by different wire lengths, detuning of the quantum-dot level, and a finite superconducting phase difference. The central result is that phase tuning converts symmetry-protected BICs into transport-active quasi-BICs and, in symmetric configurations, produces quadratic transmission zeros with a strongly enhanced electronic thermoelectric response.

The structure of this paper is organized as follows. Section \ref{secII} elucidates the model along with the methodology used to derive the quantities of interest; Section\ \ref{secIII} presents the results and their subsequent discussion; and Section\ \ref{secIV} offers the concluding remarks.

%%%%%%%%%%%%%%%%%%%%%%%%%%%%%%%%%%%%%%%%%%%%%%%%%%%%%%%%%%%%%%
\section{Model and method}\label{secII}
%%%%%%%%%%%%%%%%%%%%%%%%%%%%%%%%%%%%%%%%%%%%%%%%%%%%%%%%%%%%%%

%%%%%%%%%%%%%%%%%%%%%%%%%%%%%%% FIG.1
\begin{figure}[t]
    \centering
    \includegraphics[width=1.00\linewidth]{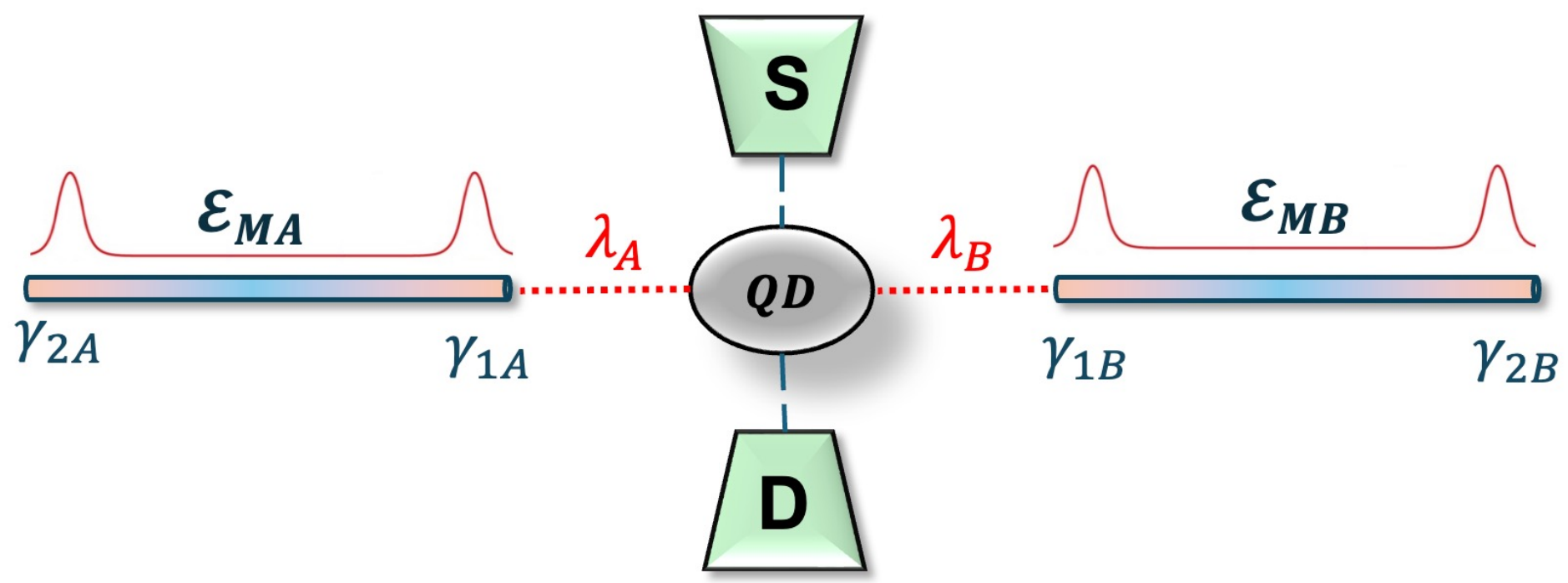}
    \caption{Schematic of the crossbar-shaped QD--TSCN device. A central quantum dot (gray) is coupled to two normal leads, source S and drain D (light green), and to two topological-superconductor nanowires $A$ and $B$ (blue and orange). Each wire hosts two MZMs, $\gamma_{1A(B)}$ and $\gamma_{2A(B)}$, whose finite-length hybridization energy is $\varepsilon_{MA(B)}$.}
    \label{fig1}
\end{figure}
%%%%%%%%%%%%%%%%%%%%%%%%%%%%%%%

We consider a crossbar-shaped QD--TSCN system in which a central QD is coupled to two normal leads and two TSCNs, each hosting one MZM at either end, as shown schematically in Fig.~ \ref{fig1}. We model the system with an effective low-energy Hamiltonian, which has the form
\begin{eqnarray}
    H &=& H_{\text{leads}} + H_\text{dots} + H_\text{dot-leads}\nonumber\\
    &+& H_\text{dot-M} + H_\text{M}\,,\label{eq1}
\end{eqnarray}where the first three terms on the right side correspond to the regular electronic contribution, given by

\begin{eqnarray}
H_{\text{leads}} &=& \sum_{\alpha,k}{ \varepsilon_{{\alpha, k}} c_{{\alpha, k}}^\dagger c_{{\alpha, k}} }\,,
\end{eqnarray}
\begin{eqnarray}
H_\text{dots} &=&  \varepsilon_{{0}} d^\dagger d,\label{Hdotfiga}
\end{eqnarray}
\begin{eqnarray}
H_{\text{dot-leads}} &=& \sum_{\alpha,k}{ V_{\alpha} d^\dag c_{{\alpha, k}}} + \text{H.c.} \,,
\end{eqnarray}
 where $c_{{\alpha, k}}^\dag ( c_{{\alpha, k}} )$ is the electron creation (annihilation) operator with momentum $k$ and energy $\varepsilon_{{\alpha, k}}$ in the lead $\alpha = \text{S,D}$. $d^\dagger (d)$ is the electron creation (annihilation) operator in the QD, with single energy level $\varepsilon_{{0}}$. The amplitude $V_{\alpha}$ describes tunneling between lead $\alpha$ and the QD.

The last two terms, $H_{\text{M}}$ and $H_{\text{dot-M}}$, describe the MZMs and their coupling to the QD, respectively. They are given by

\begin{eqnarray}
H_\text{dot-M} &=& \left( \lambda_{_{A}} d - \lambda_{_{A}}^* d{^\dagger} \right) \gamma_{_{1A}} \nonumber\\ &+& \left( \lambda_{_{B}} d - \lambda_{_{B}}^* d{^\dagger} \right) \gamma_{_{1B}} ,\label{HdotM1dot}
\end{eqnarray}
\begin{eqnarray}
H_\text{M} &=& i\varepsilon_{_{MA}}\gamma_{_{1A}}\gamma_{_{2A}} + i\varepsilon_{_{MB}}\gamma_{_{1B}}\gamma_{_{2B}}\label{HM} \text{,}
\end{eqnarray}
where $\gamma_{_{\beta l}}$ denotes the MZM operator, which satisfies both $\gamma_{_{\beta l}}=[\gamma_{_{\beta l}}]^{\dag}$ and $\{\gamma_{_{\beta l}},\gamma_{_{\beta' l'}}\}=\delta_{\beta,\beta'}\delta_{l,l'}$, being $\beta=1,2$ and $l=A,B$. The amplitude $\lambda_{A(B)}$ couples the QD to $\gamma_{1A(B)}$. The finite-length Majorana hybridization obeys $\varepsilon_{Ml}\propto e^{-L_l/\zeta}$, where $L_l$ is the wire length and  $\zeta$ is the superconducting coherence length.

A useful way to treat the system analytically is by writing each MZM as a superposition of regular fermionic operators as $\gamma_{1l}=(f_{l}+f_{l}^{\dag})/\sqrt{2}$ and $\gamma_{2l}=-i(f_{l}-f_{l}^{\dag})/\sqrt{2}$
which satisfy $\{f_{_{A(B)}},f_{_{A(B)}}\}=\{f_{_{A(B)}}^{\dag},f_{_{A(B)}}^{\dag}\}=0$ and $\{f_{_{A(B)}},f_{_{A(B)}}^{\dag}\}=1$. Without loss of generality, we choose the gauge $\lambda_A=|\lambda_A|e^{i\theta/2}$ and $\lambda_B=|\lambda_B|\in\mathbb{R}$, where $\theta$ is the superconducting phase difference between the two TSCNs. Then, Eqs.\ (\ref{HdotM1dot}), and (\ref{HM}) transform to

\begin{eqnarray}
H_\text{dot-M} &=& \frac{\left|\lambda_A\right|}{\sqrt{2}}\left(e^{i\theta/2}d -  e^{-i\theta/2}d{^\dagger} \right) \left(f_{A}+f_{A}^{\dagger}\right)\nonumber\\ &+& \frac{\left|\lambda_B\right|}{\sqrt{2}}\left( d -  d{^\dagger} \right) \left(f_{B}+f_{B}^{\dagger}\right)\text{,}
\label{HdotMfb}
\end{eqnarray}
\begin{eqnarray}
H_\text{M} &=& \varepsilon_{_{MA}}\left(f_{A}^{\dagger}f_{A}-\frac{1}{2}\right) + \varepsilon_{_{MB}}\left(f_{B}^{\dagger}f_{B}-\frac{1}{2}\right)\label{HMf}\,.
\end{eqnarray}

In the wide-band approximation, the electron and hole self-energies generated by normal lead $\alpha$ are energy independent, $\Sigma_{\alpha}^{e(h),r}=-i\Gamma_{\alpha}$, with $\Gamma_S+\Gamma_D=\Gamma$. The transport reduction to a scalar transmission is justified below for symmetric couplings, $\Gamma_S=\Gamma_D=\Gamma/2$, and an antisymmetric source--drain bias. We use the spectral-function convention $A_j(\omega)=-2\,\mathrm{Im}G_{jj}^r(\omega)$ and the corresponding local density of states $\rho_j(\omega)=A_j(\omega)/(2\pi)$. Thus, the quantity plotted for the QD is $\rho_d(\omega)=-(1/\pi)\mathrm{Im}G_{ee}^r(\omega)$, whereas the Majorana panels display the spectral functions $A_{A(B)}(\omega)$.

The Hamiltonian described in Eq.(\ref{eq1}) is spinless since only electrons with one spin projection will couple to the MZMs  \cite{ruiz2015interaction}. 

\subsection{Eigenvalues analysis}
We first analyze the eigenvalues of the isolated dot--Majorana subsystem in Fig.~ \ref{fig1}. These eigenvalues locate the molecular levels. Their identification as BICs or quasi-BICs is established from their coupling to the open transport channel and from the corresponding Green-function linewidths.

For QD energy $\varepsilon_{0} = 0$, the eigenvalues can be expressed in a compact form under the assumption of symmetric coupling ($\lambda_{A(B)} = \lambda$). These are given by
\begin{eqnarray}
    \omega^{\pm}_{0} &=& 0\label{eig_1}\,,\\ 
    \nonumber\\
	2[\omega_{1}^{\pm}]^2 &=& \varepsilon^2_{_{MA}}+\varepsilon^2_{_{MB}}+4\lambda^2 \nonumber\\ &-& \sqrt{[\varepsilon^2_{_{MA}}-\varepsilon^2_{_{MB}}]^2+8\lambda^4(1+\cos(\theta))}\label{eig_2}\,,\\
    \nonumber\\
	2[\omega_{2}^{\pm}]^2 &=& \varepsilon^2_{_{MA}}+\varepsilon^2_{_{MB}}+4\lambda^2 \nonumber\\ &+& \sqrt{[\varepsilon^2_{_{MA}}-\varepsilon^2_{_{MB}}]^2+8\lambda^4(1+\cos(\theta))}
    \label{eig_3}\,,
\end{eqnarray}
where $\omega_{0}$ has double degeneracy. \\

\subsection{Thermoelectric quantities}
Because the dot--Majorana coupling contains anomalous terms, we formulate transport in the Nambu basis $(d,d^{\dagger})$,
\begin{equation}
\mathcal{G}_d^r(\omega)=
\begin{pmatrix}
G_{ee}^r(\omega) & G_{eh}^r(\omega)\\
G_{he}^r(\omega) & G_{hh}^r(\omega)
\end{pmatrix}.
\end{equation} 

With the convention $\Sigma_\alpha^r=-i\Gamma_\alpha$, the electron-transfer and crossed-Andreev probabilities are
\begin{align}
\mathcal{T}_{\mathrm{ET}}(\omega)&=4\Gamma_S\Gamma_D|G_{ee}^r(\omega)|^2,\\
\mathcal{T}_{\mathrm{CAR}}(\omega)&=4\Gamma_S\Gamma_D|G_{he}^r(\omega)|^2.
\end{align}

We consider symmetric lead couplings, $\Gamma_S=\Gamma_D=\Gamma/2$, and the antisymmetric source--drain protocol
\begin{equation}
V_{S/D}=\pm\frac{\Delta V}{2},\,\,{\rm and}\,\,  
T_{S/D}=T\pm\frac{\Delta T}{2}.
\end{equation}

For this differential mode, the normal-lead currents satisfy $I_S=-I_D$, and electron-transfer and Andreev-conversion processes combine into
\begin{equation}
\mathcal{T}(\omega)=4\Gamma_S\Gamma_D
\left(|G_{ee}^r(\omega)|^2+|G_{eh}^r(\omega)|^2\right).
\label{Teff}
\end{equation}

The spectral identity
\begin{equation}
-\mathrm{Im}G_{ee}^r(\omega)=\Gamma
\left(|G_{ee}^r(\omega)|^2+|G_{eh}^r(\omega)|^2\right)
\end{equation}

follows from the general nonequilibrium Green's-function relation $A(\omega)=G^r(\omega)\Gamma G^{r\dagger}(\omega)$ applied to the electron row of the Nambu Green's function, i.e., from unitarity of the underlying Bogoliubov--de Gennes scattering problem \cite{datta1997electronic}, and then gives
\begin{equation}
\mathcal{T}(\omega)=-\Gamma\,\mathrm{Im}G_{ee}^r(\omega).
\label{Tspectral}
\end{equation}
Equation~\eqref{Tspectral} is therefore valid for the symmetric differential protocol specified above; for one-sided bias or asymmetric lead couplings, the normal and Andreev channels must be retained separately.

We take $e>0$ to denote the magnitude of the electron charge. With the sign convention used here, the linear-response currents are
\begin{align}
I_{\mathrm{charge}}&=-e^2\mathcal{L}_0\Delta V+\frac{e}{T}\mathcal{L}_1\Delta T,\label{icharge}\\
I_{\mathrm{heat}}&=e\mathcal{L}_1\Delta V-\frac{1}{T}\mathcal{L}_2\Delta T,\label{iheat}
\end{align}
where the exact transport moments are
\begin{equation}
\mathcal{L}_n(\mu,T)=\frac{1}{h}\int_{-\infty}^{+\infty}d\omega\,
\left(-\frac{\partial f(\omega,\mu)}{\partial\omega}\right)
(\omega-\mu)^n\mathcal{T}(\omega).
\label{Lnexact}
\end{equation}
Here $f(\omega,\mu)=[e^{(\omega-\mu)/(k_BT)}+1]^{-1}$. The electrical conductance, Seebeck coefficient, electronic thermal conductance, and electronic figure of merit are
\begin{align}
\mathcal{G}(\mu)&=e^2\mathcal{L}_0,\\
S(\mu)&=-\frac{1}{eT}\frac{\mathcal{L}_1}{\mathcal{L}_0},\\
\kappa_{\mathrm{el}}(\mu)&=\frac{1}{T}\left(\mathcal{L}_2-\frac{\mathcal{L}_1^2}{\mathcal{L}_0}\right),\label{kappamu}\\
ZT_{\mathrm{el}}&=\frac{S^2\mathcal{G}T}{\kappa_{\mathrm{el}}}
=\frac{\mathcal{L}_1^2}{\mathcal{L}_0\mathcal{L}_2-\mathcal{L}_1^2}.\label{ZT}
\end{align}
Only the electronic thermal conductance is included. A finite phonon contribution would reduce the total figure of merit according to $ZT=S^2\mathcal{G}T/(\kappa_{\mathrm{el}}+\kappa_{\mathrm{ph}})$. 

For stable numerical evaluation, Eq.~\eqref{Lnexact} is written with $x=(\omega-\mu)/(k_BT)$ as
\begin{equation}
\mathcal{L}_n^{\mathrm{ex}}=\frac{(k_BT)^n}{h}
\int_{-\infty}^{+\infty}\frac{dx}{4\cosh^2(x/2)}\,
 x^n\mathcal{T}(\mu+k_BTx).
\label{Lnscaled}
\end{equation} 
\DIFdelbegin %DIFDELCMD < 

The Sommerfeld expansion through fourth order is retained as a low-temperature analytic approximation,
\begin{align}
\mathcal{L}_0&=\frac{1}{h}\left[\mathcal{T}^{(0)}+\frac{\pi^2}{6}\mathcal{T}^{(2)}\xi^2+\frac{7\pi^4}{360}\mathcal{T}^{(4)}\xi^4+\mathcal{O}(\xi^6)\right],\label{eq18}\\
\mathcal{L}_1&=\frac{1}{h}\left[\frac{\pi^2}{3}\mathcal{T}^{(1)}\xi^2+\frac{7\pi^4}{90}\mathcal{T}^{(3)}\xi^4+\mathcal{O}(\xi^6)\right],\label{eq19}\\
\mathcal{L}_2&=\frac{1}{h}\left[\frac{\pi^2}{3}\mathcal{T}^{(0)}\xi^2+\frac{7\pi^4}{30}\mathcal{T}^{(2)}\xi^4+\mathcal{O}(\xi^6)\right],\label{eq20}
\end{align}
where $\mathcal{T}^{(n)}(\mu)=(d^n\mathcal{T}/d\omega^n)_{\omega=\mu}$ and $\xi=k_BT$. The Wiedemann--Franz law, $\mathscr{L}=\kappa_{\mathrm{el}}/(\mathcal{G}T)=\mathscr{L}_0$ with $\mathscr{L}_0=\pi^2k_B^2/(3e^2)$, applies at low temperature when the transmission is sufficiently smooth over the thermal window. Sharp quasi-BIC resonances and transmission zeros invalidate this smooth-transmission condition.

We have compared the Sommerfeld results against Eq.~\eqref{Lnscaled}. For the parameters of Fig.~\ref{fig.6} and $k_BT/\Gamma=8.6173\times10^{-5}$, exact integration gives $ZT_{\mathrm{el}}^{\max}=0.752$ in the long-wire limit and $0.749$ for $\varepsilon_{MA}=\varepsilon_{MB}=0.05\Gamma$, while the corresponding peak Lorenz ratios are $4.186$ and $4.145$. The fourth-order Sommerfeld expansion reproduces the qualitative behavior and is accurate to a few percent for these curves, but its error increases for the narrowest structures, reaching about $6\%$ in Fig.~\ref{fig.7}(h).

%%%%%%%%%%%%%%%%%%%%%%%%%%%%%%%%%%%%%%%%%%%%%%%%%%%%%%%%%%%%%%
\section{Results}\label{secIII}
%%%%%%%%%%%%%%%%%%%%%%%%%%%%%%%%%%%%%%%%%%%%%%%%%%%%%%%%%%%%%%
% FIG 2
\begin{figure}[!t]
\centering
\includegraphics[width=1.0\linewidth]{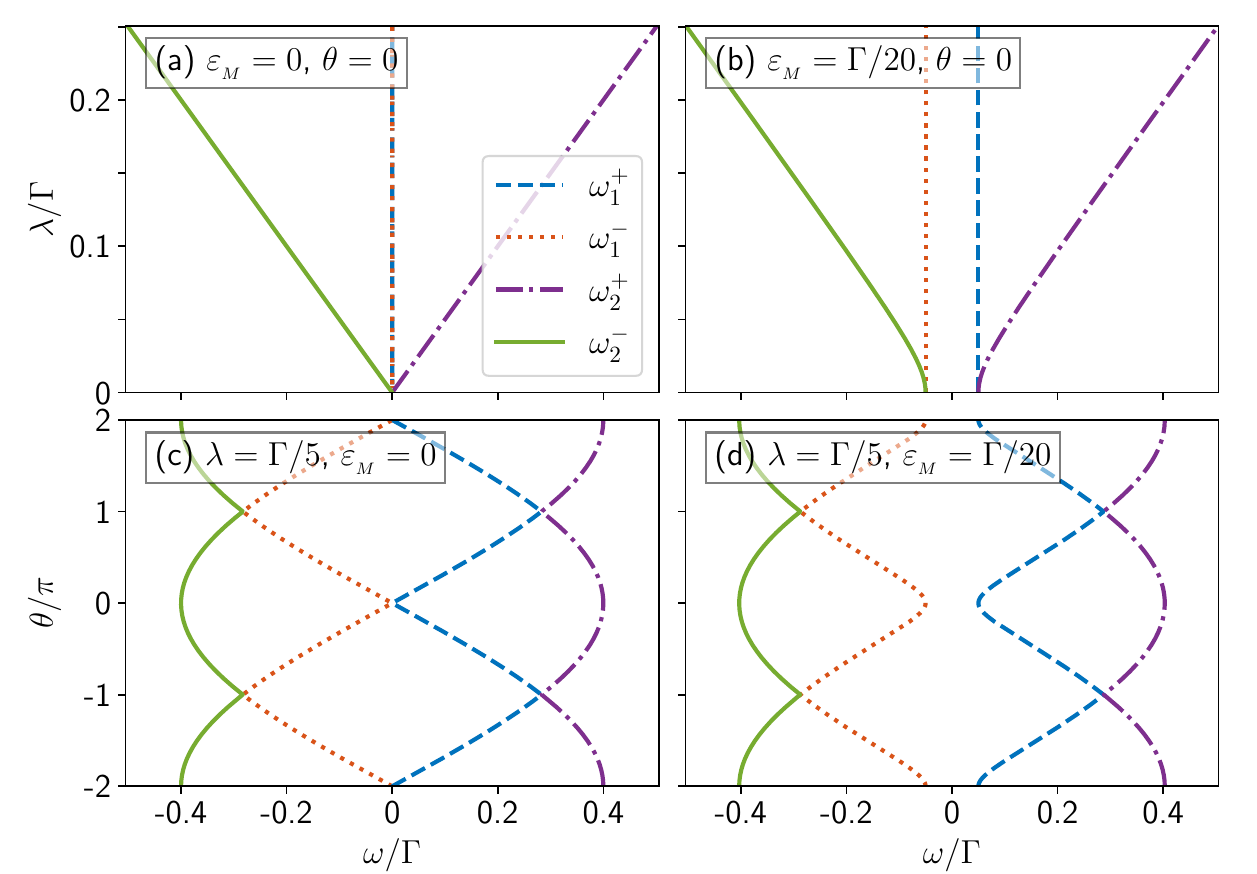}
\caption{Top panels: dependence of the eigenvalues $\omega_j^{\pm}$ on the coupling $\lambda=\lambda_{A(B)}$ at zero phase difference, $\theta=0$. Bottom panels: dependence on the superconducting phase difference for $\lambda=\Gamma/5$. The left column shows the long-wire limit, $\varepsilon_{MA(B)}=0$, and the right column the finite-wire regime, $\varepsilon_{MA(B)}=\Gamma/20$. The doubly degenerate eigenvalues $\omega_0^{\pm}=0$ from Eq.~(\ref{eig_1}) are not shown.}
\label{fig.2}
\end{figure}

%%%%%%%%%%%%%%%%%%%%%%%%%%%%%%%%%%%%%%%%%%%%%%%%%%%%%%%%%%%%%%
% FIG 3
\begin{figure}[!t]
\centering
\includegraphics[width=1.0\linewidth]{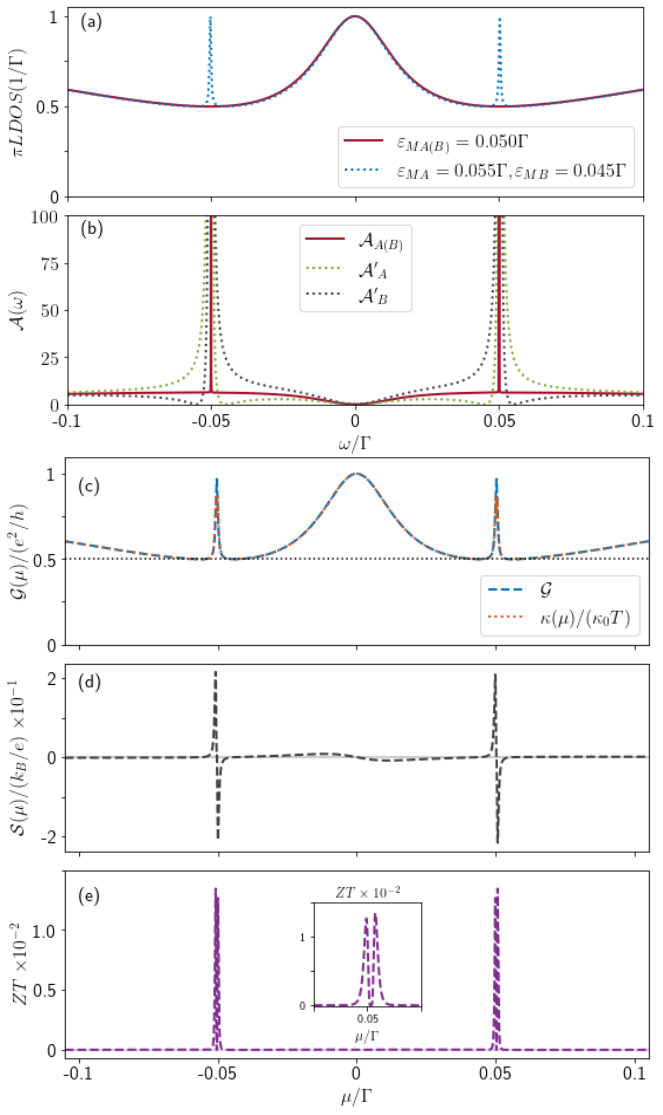}
\caption{(a) QD LDOS and (b) Majorana spectral functions versus $\omega$\; (c) electrical conductance $\mathcal{G}$ and electronic thermal conductance $\kappa_{\mathrm{el}}$; (d) Seebeck coefficient $S$; and (e) electronic figure of merit $ZT_{\mathrm{el}}$ versus  $\mu$. Wire-length asymmetry is introduced through $\varepsilon_{MA}=0.055\Gamma$ and $\varepsilon_{MB}=0.045\Gamma$ , corresponding to $\Delta_0=0.005\Gamma$ around $\varepsilon_M=0.05\Gamma$. We set $\varepsilon_0=0$ , $\theta=0$, and $k_BT/\Gamma=8.6173\times10^{-5}$. The finite height of the exact-BIC peaks in panel (b) depends on the numerical broadening.}
\label{fig.3}
\end{figure}

%%%%%%%%%%%%%%%%%%%%%%%%%%%%%%%%%%%%%%%%%%%%%%%%%%%%%%%%%%%%%%
% FIG 4
\begin{figure}[!t]
\centering
\includegraphics[width=1.0\linewidth]{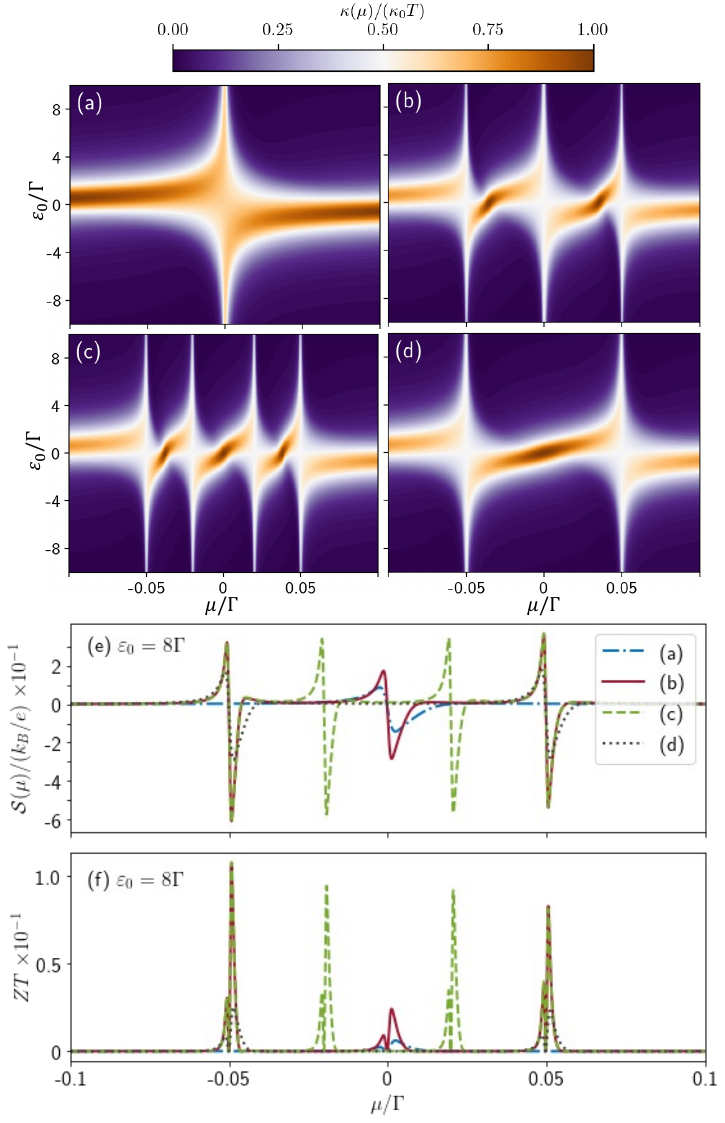}
\caption{Electronic thermal conductance $\kappa_{el}/(\kappa_0T)$ versus QD energy $\varepsilon_{0}$ and chemical potential $\mu$ for (a) $\varepsilon_{MA}=\varepsilon_{MB}=0$, (b) $\varepsilon_{MA}=0.05\Gamma$ and $\varepsilon_{MB}=0$, (c) $\varepsilon_{MA}=0.05\Gamma$ and $\varepsilon_{MB}=0.02\Gamma$, and (d) $\varepsilon_{MA}=\varepsilon_{MB}=0.05\Gamma$. Panels (e) and (f) show $S$ and $ZT_{\mathrm{el}}$, respectively at $\varepsilon_0=8\Gamma$. We set $\theta=0$ and $k_BT/\Gamma=8.6173\times10^{-5}$.}
\label{fig.4}
\end{figure}

%%%%%%%%%%%%%%%%%%%%%%%%%%%%%%%%%%%%%%%%%%%%%%%%%%%%%%%%%%%%%%
% FIG 5
\begin{figure}[!t]
\centering
\includegraphics[width=01.0\linewidth]{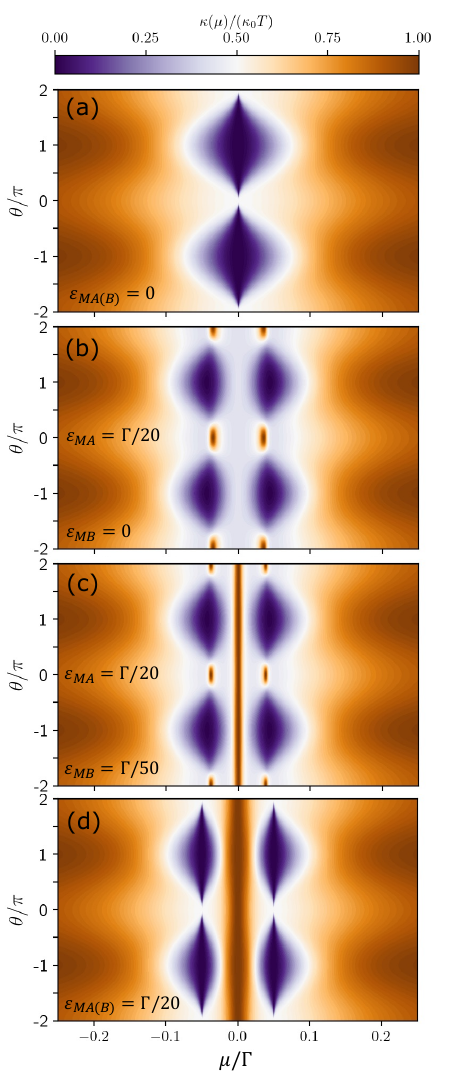}
\caption{Electronic thermal conductance $\kappa_{el}/(\kappa_0T)$ versus chemical potential $\mu$ and superconducting phase difference $\theta$ for (a) $\varepsilon_{MA}=\varepsilon_{MB}=0$, (b) $\varepsilon_{MA}=0.05\Gamma$ and $\varepsilon_{MB}=0$, (c) $\varepsilon_{MA}=0.05\Gamma$ and $\varepsilon_{MB}=0.02\Gamma$, and (d) $\varepsilon_{MA}=\varepsilon_{MB}=0.05\Gamma$. We set $\varepsilon_0=0$ and $k_BT/\Gamma=8.6173\times10^{-5}$.}
\label{fig.5}
\end{figure}
%%%%%%%%%%%%%%%%%%%%%%%%%%%%%%%%%%%%%%%%%%%%%%%%%%%%%%%%%%%%%%

%%%%%%%%%%%%  FIG 6 %%%%%%%%%%%%%%%%%%
\begin{figure}
    \centering
    \includegraphics[width=1.0\linewidth]{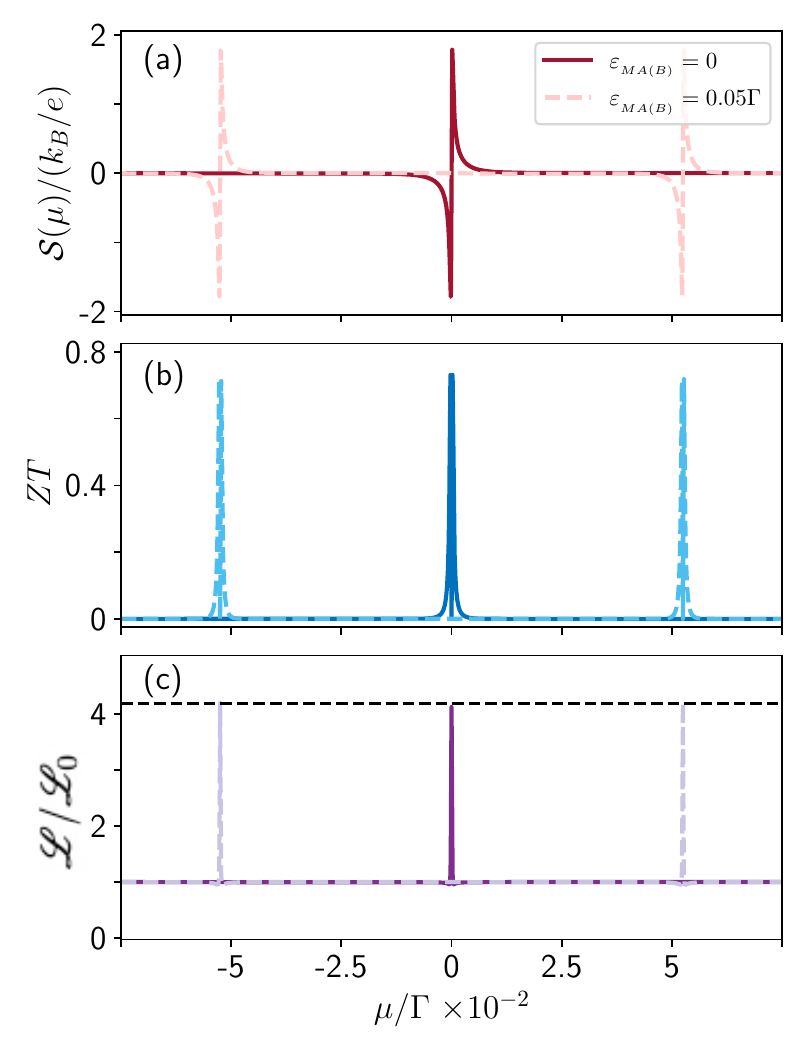}    
    \caption{(a) Seebeck coefficient $S$, (b) electronic figure of merit $ZT_{\mathrm{el}}$ , and (c) normalized Lorenz ratio $\mathscr{L}/\mathscr{L}_0$ versus $\mu$, where $\mathscr{L}_0=\pi^2k_B^2/(3e^2)$. We set $\theta=\pi/4$, $\varepsilon_0=0$, and $k_BT/\Gamma=8.6173\times10^{-5}$. Solid lines show the long-wire limit, $\varepsilon_{MA}=\varepsilon_{MB}=0$, and dashed lines the symmetric finite-wire case, $\varepsilon_{MA}=\varepsilon_{MB}=\Gamma/20$.}
    \label{fig.6}
\end{figure}
%%%%%%%%%%%%%%%%%%%%%%%%%%%%%%%

In the following, all energies are expressed in units of $\Gamma$, and the thermoelectric calculations use the dimensionless temperature $k_BT/\Gamma=8.6173\times10^{-5}$. A physical temperature thus can  be assigned only after specifying the experimental value of $\Gamma$. The low-energy Majorana description additionally requires $k_BT$, $\Gamma$, $|\lambda_{A(B)}|$, $|\varepsilon_0|$, and $\varepsilon_{MA(B)}$ to remain within the induced-gap regime.

%%%%%%%%%%%%%%explanation of figure 2:
Figure~\ref{fig.2} presents the isolated dot--Majorana eigenvalues as functions of $\lambda$ (top panels) and $\theta$ (bottom panels). In the long-wire limit at $\theta=0$, Eq.~(\ref{eig_2}) yields $\omega_1^{\pm}=0$, independent of $\lambda$, while Eq.~ (\ref{eig_3}) gives $\omega_2^{\pm}=\pm2\lambda$. For $\varepsilon_{MA}=\varepsilon_{MB}=\varepsilon_M=\Gamma/20$, the finite overlap shifts the lower pair to $\omega_1^{\pm}=\pm\varepsilon_M$ at $\lambda=0$ and gives $\omega_2^{\pm}=\pm\sqrt{\varepsilon_M^2+4\lambda^2}$. At fixed $\lambda=\Gamma/5$, the spectrum is $2\pi$ periodic in $\theta$, and at $\theta=\pm\pi$ the finite-wire branches become degenerate at $\omega_{1,2}^{\pm}=\pm\sqrt{\varepsilon_M^2+2\lambda^2}$. For symmetric wires and $\theta=2n\pi$, one superposition decouples from the dot--lead channel at $\omega=\pm\varepsilon_M$ and forms a BIC, while the orthogonal combination remains transport active. In the following we set $\lambda_A=\lambda_B=0.2\Gamma$.
%%%%%%%%%%%%%%%%%%%%%%%%%%%%%%%%%%%%%%%%%%%%%%%%%%%%%%%%%%%%%%

%%%%%%%%%% Explanation figure 3
Figure~\ref{fig.3} examines wire-length asymmetry, $\varepsilon_{MA}=\varepsilon_M+\Delta_0$ and $\varepsilon_{MB}=\varepsilon_M-\Delta_0$, with $\varepsilon_M=0.05\Gamma$ and $\Delta_0=0.005\Gamma$. In the symmetric configuration,  the wire spectral functions contain zero-linewidth BIC contributions at $\omega=\pm0.05\Gamma$, whereas these states are absent from the smooth QD LDOS. Unequal hybridization energies give the formerly dark states finite linewidths and convert them into quasi-BICs, producing narrow QD resonances near the same energies. The corresponding electrical and electronic thermal conductances display closely related lateral resonances. The Seebeck coefficient changes sign across each quasi-BIC; for the symmetric case it is negligible on the scale shown and vanishes exactly at particle--hole symmetry. Exact energy integration gives $|S|_{\max}\simeq0.20\,k_B/e$ and $ZT_{\rm el}^{\max}\simeq1.2\times10^{-2}$. The $ZT_{\rm el}$ maxima form doublets on the flanks of the quasi-BIC resonances rather than occurring exactly at $\mu=\pm0.05\Gamma$. These results show that wire-length asymmetry generates a finite but modest thermoelectric response.
%%%%%%%%%%%%%%%%%%%%%%%%%%%%%%%%%%%%%%%%%%%%%%%%%%%%%

%%%%%%%%%% Explanation figure 4:

Figure~\ref{fig.4} shows the effect of detuning the QD level. Panels (a)--(d) plot $\kappa_{\rm el}(\mu,\varepsilon_0)/(\kappa_0T)$ for the four Majorana-overlap configurations specified in the caption. The Fano structures are associated with the energies $\mu=\pm\varepsilon_{MA}$ and $\mu=\pm\varepsilon_{MB}$ and become increasingly narrow for large $|\varepsilon_0|$. Panels (e) and (f), evaluated at $\varepsilon_0=8\Gamma$, show thermoelectric doublets on the flanks of these structures. Exact integration gives $ZT_{\rm el}^{\max}\simeq0.102$ for $\varepsilon_{MA}=0.05\Gamma$, $\varepsilon_{MB}=0$, and $ZT_{\rm el}^{\max}\simeq0.105$ for $\varepsilon_{MA}=0.05\Gamma$, $\varepsilon_{MB}=0.02\Gamma$. Thus, dot detuning enhances the maximum response by approximately a factor of nine relative to Fig.~\ref{fig.3}. Away from these doublets, $ZT_{\rm el}$ is small but not identically zero. Because $\varepsilon_0=8\Gamma$ is far from resonance, a quantitative experimental mapping must ensure that this detuning remains within the validity window of the low-energy Majorana model.
%%%%%%%%%%%%%%%%%%%%%%%%%%%%%%%%%%%%%%%%%%%%%%%%%%%%%

%%%%%%%%%%%%%% Explanation figure 5:
Figure~\ref{fig.5} displays $\kappa_{\rm el}(\mu,\theta)/(\kappa_0T)$ at $\varepsilon_0=0$. In panel (a), the long-wire limit gives $\kappa_{\rm el}(0)/(\kappa_0T)\to1/2$ at $\theta=2n\pi$ and a phase-induced suppression for $\theta\neq2n\pi$, with the widest antiresonant region at $\theta=(2n-1)\pi$. In panel (b), the zero mode of wire $B$ preserves a half-conductance feature at $\mu=0$, while the lateral resonance occurs near $\mu=\pm0.035\Gamma$. Panel (c) contains corresponding structures near $\mu=\pm0.04\Gamma$ and a narrow unit-conductance region at $\mu=0$. In the asymmetric panels (b) and (c), phase tuning produces deep but generally nonvanishing antiresonant dips. In the symmetric finite-wire case, panel (d), the conductance at $\theta=2n\pi$ and $\mu=\pm0.05\Gamma$ is approximately one-half because one Majorana superposition forms a BIC while the orthogonal combination remains conducting. For $\theta\neq2n\pi$, these energies evolve into exact transmission zeros in the $T\to0$ limit and into strongly suppressed finite-temperature minima. This BIC-to-quasi-BIC conversion underlies the phase-controlled Ghost-Fano Majorana response \cite{de2003ghost,ramos2019fano}.
%%%%%%%%%%%%%%%%%%%%%%%%%%%%%%%%%%%%%%%%%%%%%%%%%%%%%

%%%%%%%%%%%%%%%Explanation of figure 6:
Figure~\ref{fig.6} displays $S$, $ZT_{\rm el}$, and the normalized Lorenz ratio $\mathscr{L}/\mathscr{L}_0$ at $\theta=\pi/4$ and $\varepsilon_0=0$. The phase difference generates quadratic transmission zeros at $\omega_0=0$ in the long-wire limit and at $\omega_0=\pm\varepsilon_M$ for equal finite overlaps. Writing $\mathcal{T}(\omega)\simeq C(\omega-\omega_0)^2$ and $x=(\mu-\omega_0)/(k_BT)$ yields
\begin{align}
\frac{S}{k_B/e}&=-\frac{2\pi^2x}{3x^2+\pi^2},\\
ZT_{\rm el}&=\frac{20\pi^2x^2}{15x^4+6\pi^2x^2+7\pi^4},\\
\frac{\mathscr{L}}{\mathscr{L}_0}&=\frac{3(15x^4+6\pi^2x^2+7\pi^4)}{5(3x^2+\pi^2)^2}.
\end{align}
Consequently, $|S|_{\max}=\pi(k_B/e)/\sqrt{3}\simeq1.81\,k_B/e$ and $ZT_{\rm el}^{\max}=0.75489$, while exactly at the transmission zero $\mathscr{L}/\mathscr{L}_0=21/5=4.2$. The Seebeck coefficient and $ZT_{\rm el}$ vanish exactly at $\mu=\omega_0$ and attain their extrema on its thermally resolved flanks; in particular, the $ZT_{\rm el}$ maxima occur at $|\mu-\omega_0|=2.5966\,k_BT$. The enhancement therefore originates from strong energy filtering and particle--hole asymmetry within the thermal window. This power-law energy-filtering mechanism is a specific realization of the general quantum-thermoelectric bounds established for scattering problems with transmission zeros \cite{whitney2014most,whitney2015finding}. At the quadratic zero, $\mathcal{L}_1=0$, $\mathcal{L}_0\propto T^2$, and $\mathcal{L}_2\propto T^4$, giving the universal Wiedemann--Franz-law violation $\mathscr{L}/\mathscr{L}_0=21/5$. Exact integration gives $ZT_{\rm el}^{\max}\simeq0.752$ for the long-wire curve and $0.749$ for the finite-wire curve. Relative to Figs.~\ref{fig.3} and \ref{fig.4}, phase tuning enhances the maximum electronic figure of merit by approximately sixtyfold and sevenfold, respectively.
%%%%%%%%%%%%%%%%%%%%%%%%%%%%%

%%%%%%%%%%%%%%%%%%%%%%%%%%%%%%%%%%%%%%%%%%%%%%%%%%%%%%%%%%%%%%
\section{Summary}\label{secIV}
%%%%%%%%%%%%%%%%%%%%%%%%%%%%%%%%%%%%%%%%%%%%%%%%%%%%%%%%%%%%%%

We have investigated thermoelectric transport through a crossbar-shaped QD coupled to two topological-superconductor nanowires hosting MZMs. Symmetry-protected BICs become finite-linewidth quasi-BICs when the nanowires have unequal lengths or a finite superconducting phase difference is introduced, thereby becoming visible in electrical and thermal transport. Unequal Majorana overlaps produce a modest response, $ZT_{\rm el}^{\max}\simeq0.012$, while strong QD detuning increases the maximum to approximately $0.105$. The largest response is generated by a superconducting phase difference, which creates quadratic transmission zeros and yields $ZT_{\rm el}^{\max}\simeq0.75$ together with the universal low-temperature Lorenz ratio $\mathscr{L}/\mathscr{L}_0=21/5$. The phase-controlled enhancement is therefore approximately sixtyfold relative to wire-length asymmetry and sevenfold relative to QD detuning. Because only the electronic thermal conductance is included, these values should be interpreted as $ZT_{\rm el}$; phonon heat transport would reduce the total figure of merit. More broadly, the results provide phase-tunable thermoelectric signatures of the Majorana-coupled interference structure and establish a route to controlling electronic energy filtering in topological hybrid nanostructures.

%%%%%%%%%%%%%%%%%%%%%%%%%%%%%%%%%%%%%%%%%%%%%%%%%%%%%%%%%%%%%%

\begin{acknowledgments}
A.P.G. is grateful for the financial support of FONDECYT Postdoctorado grant No. 3260695. J.P.R.-A is grateful for the financial support of FONDECYT Iniciación grant No. 11240637. V.J. and P.A.O acknowledge financial support from FONDECYT Grant No. 1230933.
\end{acknowledgments}
%%%%%%%%%%%%%%%%%%%%%%%%%%%%%%%%%%%%%%%%%%%%%%%%%%%%%%%%%%%%%%

\section*{DATA AVAILABILITY STATEMENT}

Data will be made available on reasonable request.

%%%%%%%%%%%%%%%%%%%%%%%%%%%%%%%%%%%%%%%%%%%%%%%
\section*{CONFLICTS OF INTEREST}

The authors declare that they have no conflict of interest.

%%%%%%%%%%%%%%%%%%%%%%%%%%%%%%%%%%%%%%%%%%%%%%%
\section*{AUTHOR CONTRIBUTION STATEMENT}

All authors contributed equally and significantly in writing this article. All authors read and approved the final manuscript.

\onecolumngrid
%%%%%%%%%%%%%%%%%%%%%%%%%%%%%%%%%%%%%%%%%%%%%%%%%%%%%%%%%%%%%%
\appendix
%%%%%%%%%%%%%%%%%%%%%%%%%%%%%%%%%%%%%%%%%%%%%%%%%%%%%%%%%%%%%%
\section{Green function}\label{AppendixA}
%%%%%%%%%%%%  FIG 7 %%%%%%%%%%%%%%%%%%
\begin{figure*}[!t]
    \centering
    \includegraphics[width=1.0\linewidth]{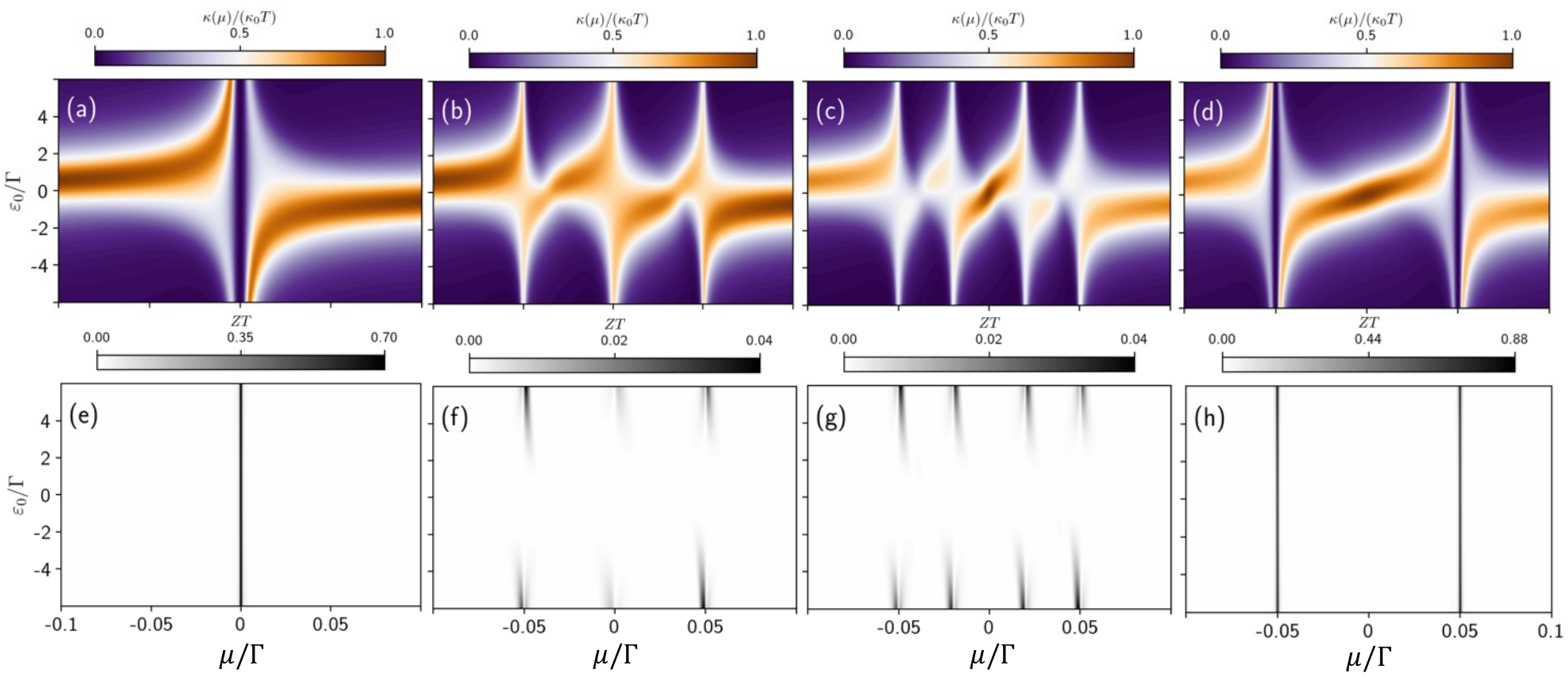}

    \caption{Top panels: electronic thermal conductance $\kappa_{\rm el}/(\kappa_0T)$ versus QD energy $\varepsilon_0$ and chemical potential $\mu$, where orange (purple) denotes maximum (minimum) conductance. We consider (a) $\varepsilon_{MA}=\varepsilon_{MB}=0$, (b) $\varepsilon_{MA}=0.05\Gamma$ and $\varepsilon_{MB}=0$, (c) $\varepsilon_{MA}=0.05\Gamma$ and $\varepsilon_{MB}=0.02\Gamma$, and (d) $\varepsilon_{MA}=\varepsilon_{MB}=0.05\Gamma$. Bottom panels (e)--(h): $ZT_{\rm el}$ for the corresponding cases. We set $\theta=\pi/4$ and $k_BT/\Gamma=8.6173\times10^{-5}$. Exact integration gives a maximum $ZT_{\rm el}\simeq0.83$ in panel (h), compared with the $O(T^4)$ Sommerfeld value $\simeq0.88$ displayed by the original color scale.}
    \label{fig.7}
\end{figure*}
%%%%%%%%%%%%  end FIG 7 %%%%%%%%%%%%%%%%%%
%%%%%%%%%%%%%%%%%%%%%%%%%%%%%%%%%%%%%%%%%%%%%%%%%%%%%%%%%%%%%%
In the Nambu basis $\Psi=(f_A,f_A^{\dagger},d,d^{\dagger},f_B,f_B^{\dagger})^T$, the full retarded Green-function takes the matrix form

\begin{widetext}
\begin{equation}
\mathbf{G}^r =
\left(
%\tiny{
%\begin{smallmatrix}
\begin{matrix} 
  	g_{_{MA}}(\omega)^{-1}                                       &  0                                                      &  \frac{\left|\lambda_{_{A}}\right|}{\sqrt{2}}e^{i\theta/2}  &  -\frac{\left|\lambda_{_{A}}\right|}{\sqrt{2}}e^{-i\theta/2}  &  0                                          &  0                                         \\
     & & & & & \\
  	0                                                      &  \widetilde{g}_{_{MA}}(\omega)^{-1}                           &  \frac{\left|\lambda_{_{A}}\right|}{\sqrt{2}}e^{i\theta/2}  &  -\frac{\left|\lambda_{_{A}}\right|}{\sqrt{2}}e^{-i\theta/2}  &  0                                          &  0                                         \\
     & & & & & \\
  	\frac{\left|\lambda_{_{A}}\right|}{\sqrt{2}}e^{-i\theta/2}  &  \frac{\left|\lambda_{_{A}}\right|}{\sqrt{2}}e^{-i\theta/2}  &  g_{0}(\omega)^{-1}                                        &  0                                                       &   \frac{\left|\lambda_{_{B}}\right|}{\sqrt{2}}   &   \frac{\left|\lambda_{_{B}}\right|}{\sqrt{2}}  \\
     & & & & & \\
  	-\frac{\left|\lambda_{_{A}}\right|}{\sqrt{2}}e^{i\theta/2}  &  -\frac{\left|\lambda_{_{A}}\right|}{\sqrt{2}}e^{i\theta/2}  &  0                                                     &  \widetilde{g}_{_{0}}(\omega)^{-1}                              &  -\frac{\left|\lambda_{_{B}}\right|}{\sqrt{2}}   &  -\frac{\left|\lambda_{_{B}}\right|}{\sqrt{2}}  \\
     & & & & & \\
  	0                                                      &  0                                                      &  \frac{\left|\lambda_{_{B}}\right|}{\sqrt{2}}               &  -\frac{\left|\lambda_{_{B}}\right|}{\sqrt{2}}                &  g_{_{MB}}(\omega)^{-1}                           &  0                                         \\
     & & & & & \\
  	0                                                      &  0                                                      &  \frac{\left|\lambda_{_{B}}\right|}{\sqrt{2}}               &  -\frac{\left|\lambda_{_{B}}\right|}{\sqrt{2}}                &  0                                          &  \widetilde{g}_{_{MB}}(\omega)^{-1}
\end{matrix}
%}
\right)^{-1}
\text{,}
\label{matrix}
\end{equation}
\end{widetext}

where the diagonal matrix elements are given by
\begin{eqnarray}
	g_{0}(\omega)^{-1}                    &=& \omega -\varepsilon_{0}+i\Gamma\text{,}\nonumber\\
	\widetilde{g}_{0}(\omega)^{-1}        &=& \omega +\varepsilon_{0}+i\Gamma\text{,}\nonumber\\
	g_{_{MA(B)}}(\omega)^{-1}             &=& \omega -\varepsilon_{_{MA(B)}}+i0^+\text{,}\nonumber\\
	\widetilde{g}_{_{MA(B)}}(\omega)^{-1} &=& \omega +\varepsilon_{_{MA(B)}}+i0^+\text{,}
\end{eqnarray}

Here $0^+$ denotes the retarded infinitesimal. In numerical spectral plots it is replaced by a small finite broadening $\eta$; consequently, the plotted height of an exact BIC is broadening dependent, whereas its position and vanishing physical linewidth are not. 

%%%%%%%%%%%%%%%%%%%%%%%%%%%%%%%%%%%%%%%%%%%%%%%%%%%%%%%%%%%%%%
\section{Additional thermoelectric results}\label{AppendixB}
%%%%%%%%%%%%%%%%%%%%%%%%%%%%%%%%%%%%%%%%%%%%%%%%%%%%%%%%%%%%%%

Figure~\ref{fig.7} extends the phase-biased calculation to finite QD detuning. In the symmetric configurations, panels (a) and (d), the phase-induced transmission zeros remain pinned at $\mu=0$ and $\mu=\pm0.05\Gamma$, respectively, independently of $\varepsilon_0$. Their associated high-$ZT_{\rm el}$ ridges are therefore robust in position, although their amplitudes vary moderately with QD detuning. Exact integration gives $ZT_{\rm el}^{\max}\simeq0.752$ at $\varepsilon_0=0$ and approximately $0.785$ near the edge of panel (e); for panel (h), the corresponding values increase from approximately $0.749$ to $0.83$. In the asymmetric configurations, panels (b) and (c), exact destructive interference is lost. The structures at $\mu=\pm\varepsilon_{MA}$ and $\pm\varepsilon_{MB}$ become half-conductance Fano crossings with adjacent peaks and dips, and the maximum response reaches only $ZT_{\rm el}\simeq0.04$ at large $|\varepsilon_0|$. Thus, the antiresonance energies in the symmetric cases are independent of the QD level, whereas the maximum thermoelectric amplitudes are not.
%%%%%%%%%%%%%%%%%%%%%%%%%%%%%%%%%%%%%%%%%%%%%%%%%%%%%%%%%%%%%%

\twocolumngrid

%\nocite{*}

\bibliographystyle{apsrev4-1}
\bibliography{bibprepint}% Produces the bibliography via BibTeX.

\end{document}